\documentclass[preprint,showkeys,preprintnumbers,amsmath,amssymb]{revtex4}

\usepackage{graphicx}
\usepackage{dcolumn}
\usepackage{bm}

\begin{document}


\title{Modified permittivity observed in bulk Gallium Arsenide and Gallium Phosphide samples at $50$ K using the Whispering Gallery mode method}

\author{John G. Hartnett$^1$, David Mouneyrac$^{1,2}$, Jean-Michel Le Floch$^1$, Jerzy Krupka$^3$, Michael E. Tobar$^1$, D. Cros$^2$}
 \email{john@physics.uwa.edu.au}
\affiliation{$^1$School of Physics, University of Western Australia 35 Stirling Hwy Crawley 6009 W.A. Australia \\ $^2$XLIM UMR 6172 ­ Université de Limoges/CNRS ­ 123 Avenue Albert Thomas 87060 Limoges, Cedex, France \\ $^3$Institute of Microelectronics and Optoelectronics Department of Electronics Warsaw University of Technology Warsaw, Poland}

\date{\today}

\begin{abstract}
Whispering Gallery modes in bulk cylindrical Gallium Arsenide and Gallium Phosphide samples have been examined both in darkness and under white light at $50$ K. In both samples we observed change in permittivity under light and dark conditions. This results from a change in the polarization state of the semiconductor, which is consistent with a free electron-hole creation/recombination process. The permittivity of the semiconductor is modified by free photocarriers in the surface layers of the sample which is the region sampled by Whispering Gallery modes.
\end{abstract}

\keywords{whispering gallery mode method, semiconductor permittivity, polarization}
\maketitle

The Whispering Gallery Mode (WGM) method has been used to make  the most accurate measurements of the complex permittivity of extremely low-loss dielectric materials, both ceramic and crystalline. The method has been employed for very precise measurements of the permittivity and the dielectric losses of both isotropic and uniaxial anisotropic materials.  \cite{Baker-Jarvis1998, Krupka1997, Krupka1999a}  The WGM technique has also been used to characterize the complex permittivity of semiconductors, including bulk monocrystalline Silicon, \cite{Krupka2006} Gallium Arsenide (GaAs)  and Gallium Phosphide (GaP) , \cite {Krupka2008} at microwave frequencies from cryogenic temperatures to room temperature. For a full description of the method see the aforementioned references. 

In this paper we present the measurement of the change in permittivity resulting from polarization changes in the semiconductors GaAs and GaP under white light at cryogenic temperatures.   The effect may find application in the design and construction of highly sensitive microwave phase shifters or tunable filters that can operate at cryogenic temperatures. Gallium Phosphide would be most suitable, because there was only a small degrading of the resonant mode Q-factor under illumination. And since the change in permittivity of the bulk semiconductor was observed to depend on the intensity of the light such devices would be highly tunable.  

We observed the time evolution of the frequency of Whispering Gallery (WG) modes while GaP and GaAs mono-crystals were illuminated with white light, and after the light was extinguished. We used mechanically polished cylindrical samples of pure GaAs (diameter = $25.39 \pm  0.01$ mm and height = $6.25 \pm  0.01$ mm) and GaP (diameter = $48.12 \pm  0.03$ mm and height of $5.00 \pm  0.01$ mm). \cite{Krupka2008} The internal dimensions of the cylindrical copper cavities used are as follows:  diameter = 34.6 mm and height = 25.8 mm for the GaAs sample; diameter = 60.0 mm and height = 43.0 mm for the GaP sample. 

The resonators were coupled to a vector network analyzer (VNA) via coaxial cables. They were located in an evacuated chamber and cooled on the cold-finger of a single-stage cryocooler. At 50 K, high order azimuthal mode numbered WG modes were identified in both crystals: in GaAs, the 18.949 GHz mode with an azimuthal mode number $m = 13$ and an electric energy filling factor $p_{\varepsilon} = 0.98577$, and in GaP, the 11.544 GHz mode with $m = 12$ and  $p_{\varepsilon} = 0.96672$. The mode resonance frequency, loaded Q-factor and couplings were determined from the measured VNA S-parameters. A continuous data acquisition system was implemented that quickly recorded the frequency of the resonant mode under examination.  Couplings were set sufficiently low not to perturb the mode frequency.

Each sample was illuminated with light that was conducted into the cryogenic resonator via optical fibers. The energy gap for GaAs and GaP are 1.424 eV and 2.26 eV respectively. \cite{GaAs, GaP} Therefore we expect threshold wavelengths of the illuminating light, to excite photoconduction and shift the resonance frequency, to be 871 nm (IR) and 549 nm (green) respectively. A white light source, emitting a maximum 10 mW of total light power, was used in both cases. This power was measured out of the fiber at room temperature before cooling. The source emits significant IR. Red (660 nm) 5 mW laser light induced no mode frequency shift in the GaP sample, but green (532 nm) 5 mW laser light induced a significant shift, and reduced the mode Q-factor. Resonant mode frequencies in the GaAs sample were shifted by red and green laser light as well as the white light source. There the microwave losses were so significant that the mode was almost completely lost while the light was on.

To convert fractional frequency of the measured mode to the real part of the permittivity ($\varepsilon_r$) we used the fact that 
\begin{equation} \label{eqn:fracfreq}
\frac{\Delta  f}{f} = -p_{\varepsilon}\frac{1}{2}\frac{\Delta \varepsilon_r}{\varepsilon_r},
\end{equation}
where $\Delta f = f-f_0$, $f_0$ is the initial mode frequency. It has been assumed here that any dimensional dependence may be be neglected (because the chosen WG mode has a sufficiently high enough azimuthal mode number that its frequency is not influenced by the cavity walls) and there are no paramagnetic impurities present.  

With both GaAs and GaP samples the white light source was switched on, frequency data recorded and then after the light was switched off the frequency data again recorded. Figures 1 and 2, respectively,  show $\Delta \varepsilon_r/\varepsilon_r$ derived from (\ref{eqn:fracfreq}) and the fractional frequency data of the N1-13 18.949 GHz mode (in GaAS) and the S1-12 11.544 GHz mode (in GaP), at 50 K. Figure 3 also shows $\Delta \varepsilon_r/\varepsilon_r$ in GaP at 11.544 GHz but with different total light power levels. The slow fluctuations in Fig. 1 are the result of temperature induced changes even though the cavity containing the semiconductor was temperature controlled at 50 K, the stability was not very good. 

In Fig. 2 it is clear there are four time constants involved; two after switching on the light and also two after switching it off. The overshoot is also clearly seen, where the mode frequency, hence fractional permittivity, overshoots the starting value before the light was turned on. Each portion of this data was then curve fitted with exponentials over its appropriate time periods.  See the dashed (red) curves--mostly hidden in the data. The latter employed 10 mW of white light power. Here however this data was collected after the sample had been previously illuminated with light. Because of persistent currents \cite{Queisser1979, Lang1979, Queisser1984}  the permittivity never returned to its initial value in darkness after cooling from room temperature. This is why the same value of $\varepsilon_r$ is shown initially and at the end of the data. Hence the cryocooler was switched off and the sample warmed up and recooled and the experiment repeated with 6 mW of power and then repeated again with 3 mW of power. In Fig. 3 the 'in darkness' initial value is not seen to be the same as the final value after cycling the the light on then off. 

Similarly GaAs was evaluated and is shown in Fig. 1 where the light was switched off. Here also the value of $\varepsilon_r$ does not return to its initial 'in darkness' value.  Note that we used a halogen lamp and considered the two time constants resulting from transients while switching the light on and off. To test this we tried rapidly disconnecting the optical fiber from the source, so that there was an abrupt decrease in light intensity and we still observed two decay times. Also since the second time constant, in GaP, is of the order of 2 minutes it is inconsistent with lamp transients. The two decay times could result from two different recombination mechanisms.

After switching off the white light source, $\Delta \varepsilon_r/\varepsilon_r$ in the GaP sample can be described by the following relations.
\begin{equation} \label{eqn:eps}
\left(\frac{\Delta \varepsilon_r}{\varepsilon_r}\right)^{-1} = \sum_{i=1}^2 \left (\frac{\Delta \varepsilon_r}{\varepsilon_r}|_i \right)^{-1} ,
\end{equation}
where
\begin{equation} \label{eqn:epsi}
\frac{\Delta \varepsilon_r}{\varepsilon_r}|_i =  B_i+C_i\left\{1-Exp\left[{\frac{t-t_0}{\tau_{i}}}\right]\right\}^{-1},
\end{equation}
where $t_0$ is the switching time, $i = 1,2$ indicates a different time constant $\tau_i$, and the constants $B_i$ and $C_i$ are to be determined. The equations (\ref{eqn:eps}), (\ref{eqn:epsi}) fit the data very well as shown in Fig. 2 after the light is switched off. From this fit the constants were evaluated and are listed in the second row of Table I. And by modifying (\ref{eqn:eps}) we can also fit the exponential dependence of (\ref{eqn:epsi}) to the $\Delta \varepsilon_r/\varepsilon_r$ data when the light is on. The following also fits the data in Fig. 2 very well after the light is switch on.
\begin{equation} \label{eqn:epson}
\left (\frac{\Delta \varepsilon_r}{\varepsilon_r}-D \right)^{-1} = \sum_{i=1}^2 \left (\frac{\Delta \varepsilon_r}{\varepsilon_r}|_i \right)^{-1} ,
\end{equation}
where $D = 1.8 \times 10^{-5}$ and the other constants as listed in the first row of Table I. In these measurements we  determined the fractional frequency better than a part in $10^6$ and with curve fit residuals of order unity, in Figs 1 and 2. The errors on the time constants are better than a few percent.

The frequency data for the GaAs sample (Fig. 1) is not so indicative of two time constants. So (\ref{eqn:eps}) was fitted with a single time constant to $\Delta \varepsilon_r/\varepsilon_r$ data, where the light was off. The resulting best fit constants are listed in the last row of Table I. From Fig. 1 it is evident that there is another much longer time constant but the fluctuations in frequency due to temperature instability makes it difficult to determine precisely. 

A change in permittivity results from a change in polarization of a crystal lattice. In this case, that can be understood in terms of the hole creation/annihilation process, modifying the bulk polarization of the material under test, as electrons are initially freed and then recombine with holes. In Fig. 2 the permittivity, hence polarization, actually overshoots the initial value after turning the light on and off again. This may result from an `oscillation' in the free electron number during the recombination process. No overshoot occurs in $\Delta \varepsilon_r/\varepsilon_r$ under reduced light power (see Fig. 3).

Also a second order relaxation system was observed in the permittivity of GaP. The decay times in light and darkness are not the same -- only the longer time constant was found to be the same.  In GaAs the WG mode frequency stability was inadequate to fully determine a second time constant in the permittivity relaxation. 

To our knowledge this is the first time that modified permittivity has been observed in a bulk semiconductor resulting from free photocarriers, though this has previously been observed under UV radiation in thin ferroelectric films. \cite{Alford2005} It is expected that the photocarriers are generated near the surface of the semiconductor where the light field has influence. But because the Whispering Gallery modes at high azimuthal mode numbers $m>10$ are largely confined to the boundary layer just inside the sample under test, as indicated by the electric energy filling factors, this method is relevant to this region. 

This work was supported by the Australian Research Council, and the Institute of Electronic Materials Technology, Warsaw, Poland, who supplied the samples.

\newpage

\begin{table}[ph]
\begin{center}
\caption{\label{tab:table3}$\Delta \varepsilon_r/\varepsilon_r$ data: top 2 rows GaP; bottom row GaAs.\\L= in light; D = in darkness. }
\begin{tabular}{lcccccccc}
\hline\hline
& &$B_1[\times 10^6]$&$C_1[\times 10^6]$&$\tau_1$ [s]&$B_2[\times 10^6]$&$C_2[\times 10^4]$&$\tau_2$ [s]&$t_0$ [s]  \\
\hline
&L&$-8.0$	& $1.2$	&$30$	&$-1.6$	&$4.7$	&$120$	&$610$ \\
&D&$8.0$	& $-1.2$	&$50$	&$-0.84$	&$-1.9$	&$120$	&$2450$ \\
\hline
&D&$20.5$	& $-6.2$	&$102$	&-	&-	&-	&$3384$ \\
\hline
\end{tabular}
\end{center}
\end{table}

\newpage
Figure captions

Fig 1: GaAs at 50 K: $\Delta \varepsilon_r/\varepsilon_r$ at 18.949 GHz under white light and in darkness.  A 200 s interval was allowed between switching. The gray dots are the measured data and the black dots result from smoothing the data by taking a sliding average of 20 adjacent measurements. The dashed (red) curve is the curve fit.

Fig2: GaP at 50 K: $\Delta \varepsilon_r/\varepsilon_r$ at 11.544 GHz under white light and in darkness. About a 30 minute interval was allowed between switching. Here, 10 mW of total light power was used. The gray dots are the measured data and the black dots result from smoothing the data by taking a sliding average of 5 adjacent measurements. The dashed (red) lines are curve fits, fitted only over the range of the black dots shown.

Fig 3: GaP at 50 K: $\Delta \varepsilon_r/\varepsilon_r$ at 11.544 GHz under white light and in darkness. Also a 30 minute interval was allowed between switching. The gray dots are the measured data under 3 mW of total light power and the black dots with 6 mW.

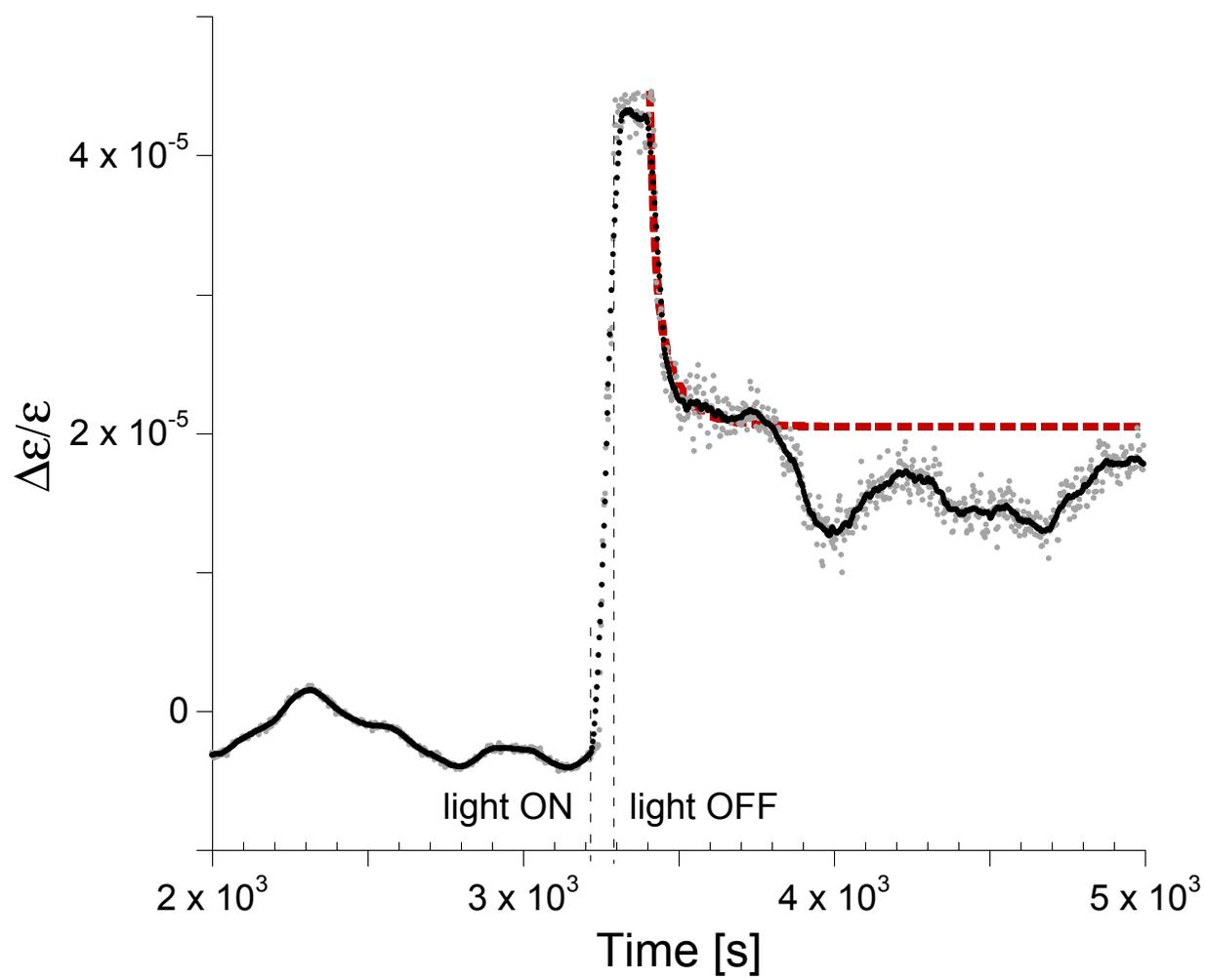
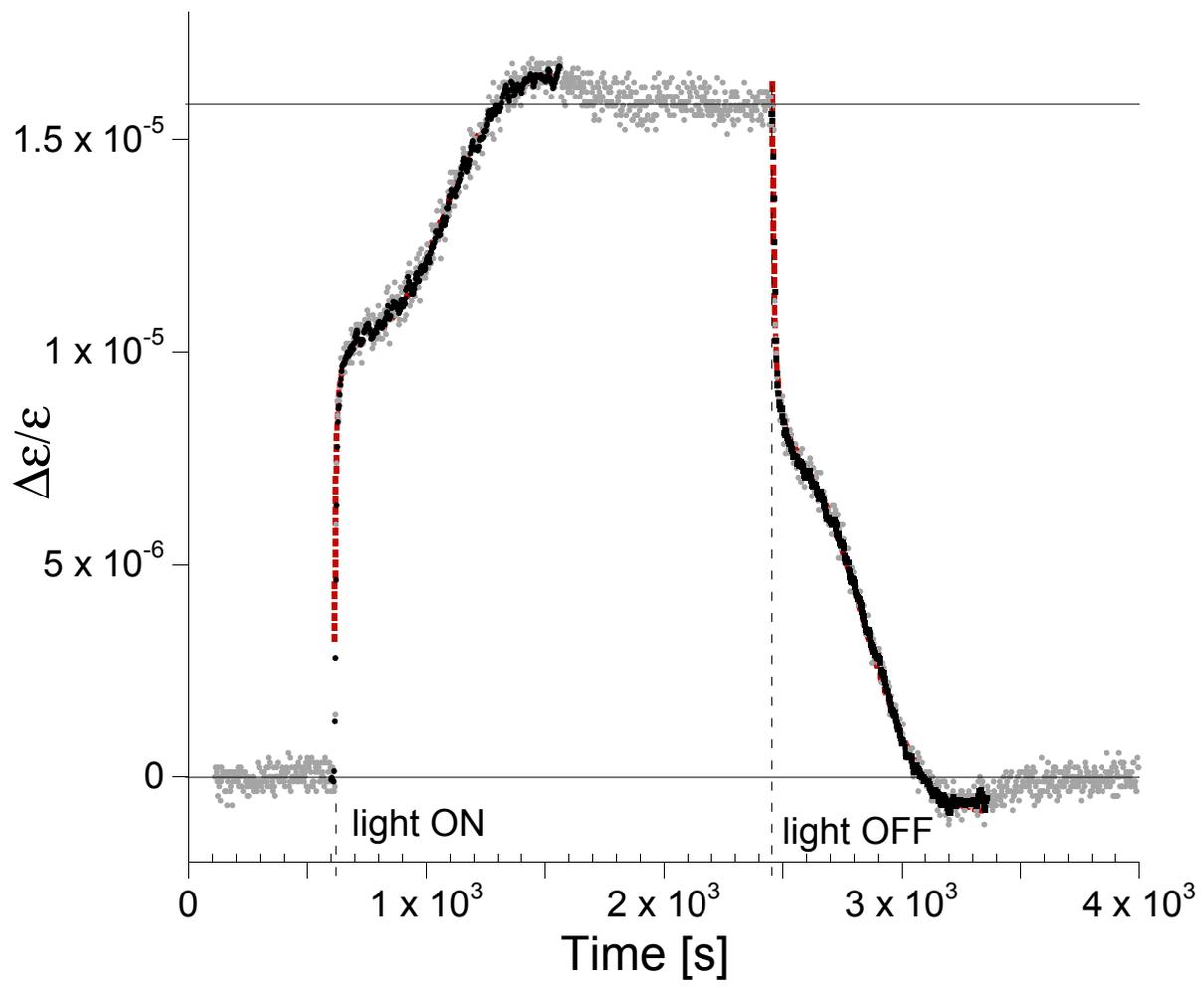
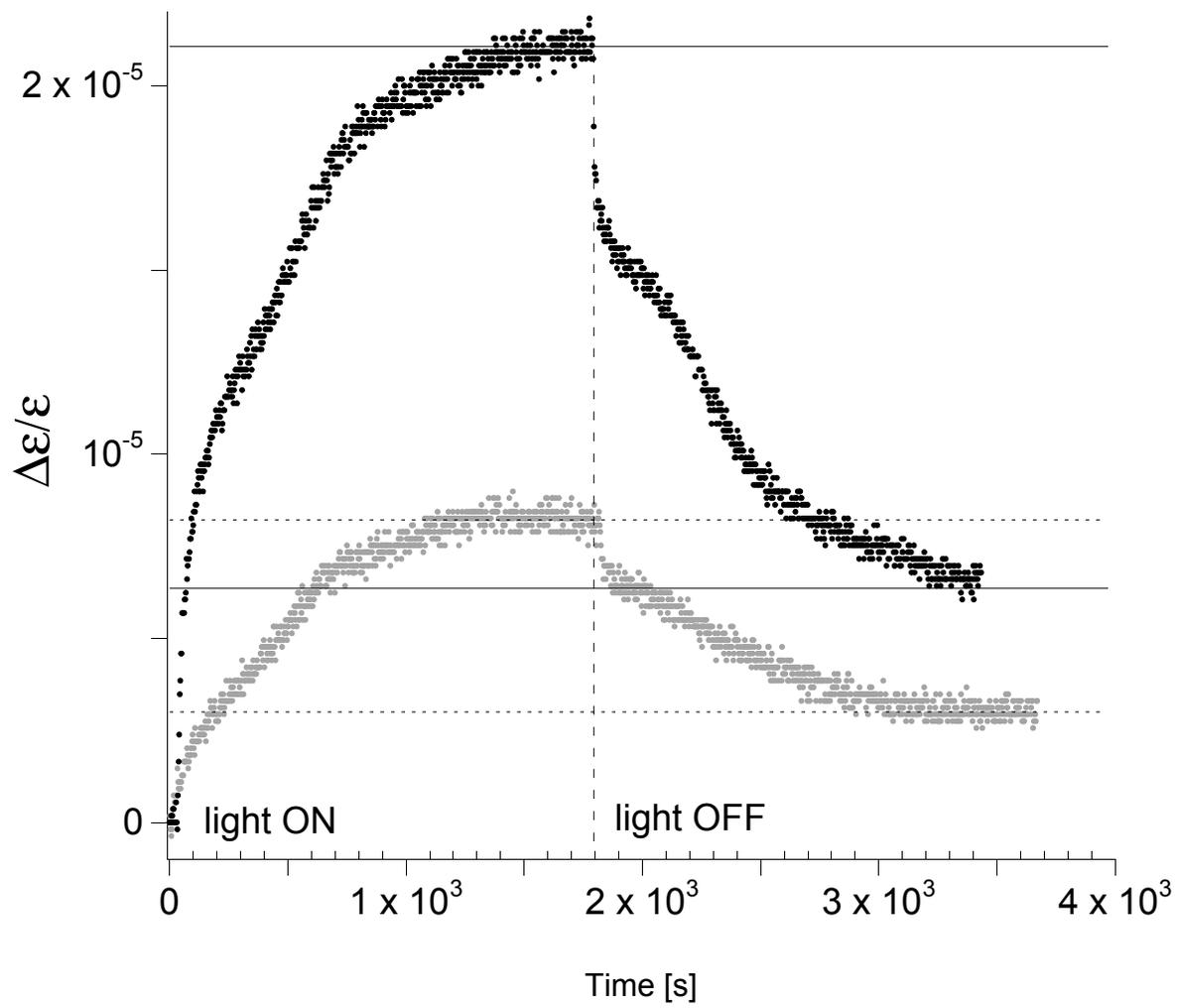
\end{document}